\documentclass[10pt, aps, prd, amsmath, floats, floatfix, twocolumn, notitlepage,
superscriptaddress, nofootinbib, showpacs]{revtex4-1}
\allowdisplaybreaks
\usepackage{float}
\usepackage[utf8]{inputenc}
\usepackage[T1]{fontenc}
\usepackage{comment}
\usepackage{bm}
\usepackage[normalem]{ulem}\usepackage{mathtools,amsmath,amssymb,amsfonts,mathrsfs,eucal,graphicx,tensor,csquotes,accents,commath,chngcntr,siunitx}
\usepackage[dvipsnames]{xcolor}
\usepackage[unicode]{hyperref}
\usepackage{orcidlink}
\hypersetup{colorlinks=true, citecolor=MidnightBlue,
            linkcolor=MidnightBlue, urlcolor=MidnightBlue, linktocpage=true}
\usepackage[normalem]{ulem}

\DeclareMathAlphabet{\mathpzc}{OT1}{pzc}{m}{it}

\definecolor{darkgreen}{rgb}{0.0, 0.6, 0.0}

\newcommand{\JHU}{William H. Miller III Department of Physics and Astronomy, Johns Hopkins University, 3400 North Charles Street, Baltimore, Maryland, 21218, USA}

\newcommand{\ICTS}{International Centre for Theoretical Sciences, Tata Institute of Fundamental Research, Bangalore 560089, India}

\newcommand{\IACS}{School of Physical Sciences, Indian Association for the Cultivation of Science, Kolkata-700032, India}

\newcommand{\IUCAA}{Inter-University Centre for Astronomy and Astrophysics, Pune 411007, India}

\newcommand{\IITGN}{Indian Institute of Technology Gandhinagar, Palaj 382055, India}

\newcommand{\WSU}{Department of Physics and Astronomy, Washington State University, 1245 Webster, Pullman, Washington 99164-2814, USA}

\labelformat{section}{Section #1} 
\labelformat{subsection}{Section \thesection.#1} 
\labelformat{subsubsection}{Section #1}
\labelformat{subsubsubsection}{Section #1}
\labelformat{equation}{Eq.~(#1)} 
\labelformat{figure}{Fig.~#1} 
\labelformat{subfigure}{Fig.~\thefigure#1} 
\labelformat{table}{Table~#1} 
\labelformat{appendix}{Appendix #1}
\begin{document}

\date{\today}

\title{Universal Ladder Structure Across Scales: From Quantum to Black Hole Physics}

\author{Rajes Ghosh\orcidlink{0000-0002-1264-938X}}
\email{rghosh13@jh.edu}
\affiliation{\JHU}
\affiliation{\ICTS}
\author{Rajendra Prasad Bhatt\orcidlink{0009-0004-9088-2998}}
\email{rajendra@iitgn.ac.in}
\affiliation{\IUCAA}
\affiliation{\IITGN}
\author{Sumanta Chakraborty\orcidlink{0000-0003-3343-3227}}
\email{tpsc@iacs.res.in}
\affiliation{\IACS}
\author{Sukanta Bose\orcidlink{0000-0002-4151-1347}}
\email{sukanta@wsu.edu}
\affiliation{\WSU}


\begin{abstract}
    Second-order ordinary linear differential equations appear ubiquitously across physics, describing the behavior of systems from the quantum world of atoms to the classical world of gravitating bodies. We present a unified symmetry-based framework that provides a ``litmus-test criterion'' to determine when such a system admits a hierarchical ladder structure, and, whenever it does, explicitly constructs the ladder. This approach uncovers a previously underappreciated connection to supersymmetric quantum mechanics and a deep commonality among diverse physical problems. Applications to the quantum harmonic oscillator and dynamical tidal response of Kerr black holes are presented to illustrate the framework. 
\end{abstract}

\maketitle

\section{Introduction}
Symmetry lies at the heart of nature and the physical laws that govern it~\cite{weyl1983symmetry}. It underpins conservation principles and organizes interactions from the quantum scale of atoms to the large-scale dynamics of gravitation~\cite{coleman1988aspects, Schwichtenberg:2015jcd}. When a system exhibits symmetries, its physical characteristics are often encoded in hidden algebraic structures that greatly simplify both their conceptual and mathematical understanding~\cite{elliott1979symmetry1, elliott1979symmetry2}. A canonical example is the quantum harmonic oscillator, whose energy spectrum emerges more elegantly from its ladder structure via the raising and lowering operators, rather than from directly solving the Schrödinger equation~\cite{Griffiths:1995, Zettili:2009}. Analogous ladder structures also arise in supersymmetric quantum mechanics (SUSY QM)~\cite{Cooper:1994eh}, eigenvalue problems and the theory of special functions (e.g., the (confluent) hypergeometric function) satisfying second-order differential equations~\cite{infeld1941new, Infeld:1951mw, U_Laha_1986}. Yet, despite their widespread applications and foundational importance, a unified understanding of why and when such ladder structures come to bear is still lacking, thereby limiting our ability to identify symmetry-based simplifications in novel and intricate physical systems.

Interestingly, the appearance of second-order differential equations is an ubiquitous feature in physics. In non-relativistic QM, every time-independent problem reduces to a similar equation through the Schrödinger framework~\cite{Griffiths:1995, Zettili:2009}. Moreover, in General Relativity (GR), scalar/vector/gravitational perturbations of Schwarzschild and Kerr black holes (BHs) likewise reduce to master second-order radial equations, whether in Klein-Gordon, Regge-Wheeler, Zerilli, or Teukolsky form~ \cite{Regge:1957td,Edelstein:1970sk,Zerilli:1970se,Teukolsky:1972my,Teukolsky:1973ha,Press:1973zz,Teukolsky:1974yv}. Even classical problems such as the scattering and stability analysis of compact objects ultimately rely on the same mathematical formalism~\cite{Vishveshwara:1970zz, Futterman:1988ni}. This ubiquity underscores some natural and important questions: Under what conditions do these apparently unrelated physical equations admit a ladder structure, and what do those symmetries reveal about the underlying physics? 

Exploring these questions is timely, especially, for ascertaining the tidal Love numbers (TLNs) of BHs. Several recent works have shown that the vanishing of static TLNs of BHs is a direct consequence of a ladder symmetry of the underlying perturbation equations~\cite{Hui:2021vcv, Charalambous:2021kcz, Hui:2022vbh, BenAchour:2022uqo, Charalambous:2022rre, Sharma:2024hlz, Rai:2024lho, Sharma:2025xii}. Yet, these case-by-case analyses, while illuminating and important, fail to address the broader questions posed above. In fact, they have largely been restricted to static perturbation equations of (confluent) hypergeometric type. Additionally, such approaches do not naturally extend to the more general dynamical perturbations encountered in realistic astrophysical settings \cite{LeTiec:2020bos, Chia:2020yla, Bhatt:2023zsy, Bhatt:2024yyz, HegadeKR:2024agt, Chakraborty:2023zed, Silvestrini:2025lbe, Chakraborty:2025wvs, Chia:2023tle, Saketh:2023bul, Katagiri:2024wbg, Combaluzier--Szteinsznaider:2025eoc}.  

In this work, we develop a unified framework that addresses these shortcomings. We derive a general existence criterion for ladder structure associated with a generic second-order ordinary linear differential equations (OLDEs), without relying on special-functions or limiting subcases. Remarkably, we find that the resulting framework has a natural and previously not-well-recognized structure akin to SUSY QM in the general case. Our construction reduces the analysis to a ``litmus-test criterion'', a necessary and sufficient condition for the existence of a ladder structure. This criterion thus provides a major conceptual simplification and offers a practical tool for identifying hidden ladder structures whenever they are present. As a result, our approach brings a range of seemingly unrelated, previously domain-specific problems under a common mathematical framework. For concreteness, we demonstrate its application for various systems, including the quantum harmonic oscillator and dynamical TLNs of Kerr BHs.

The implications of our results span multiple areas of physics. Particularly in QM, they provide an algebraic tool for identifying solvable models beyond a handful of standard textbook examples, enabling symmetry-driven construction of energy spectra and eigenfunctions~\cite{Griffiths:1995, Zettili:2009}. In gravitational physics, the framework offers a novel and unified pathway to study BH perturbation equations, yielding crucial insights in TLNs. In future, our results may facilitate the computation of BH quasi-normal modes and scattering amplitudes using ladder-operator techniques, providing more analytical handles and streamlining numerical analyses. Since TLNs affect gravitational wave phasing during compact binary inspirals, the ability to compute and classify them through symmetry-based methods can have particularly far-reaching consequences in observational tests of the BH nature of high-mass binary components and GR.

This paper is arranged as follows: In  \ref{Section: Ladder structure for a general second-order differential equation}, we discuss the ladder structure of a general second-order ordinary linear differential equation. In \ref{Section: Applications}, we provide applications of the formalism developed in \ref{Section: Ladder structure for a general second-order differential equation} for a quantum harmonic oscillator (\ref{Section: Spectrum of quantum harmonic oscillator}) and dynamical TLNs of Kerr BHs (\ref{Section: Dynamical tidal response of Kerr BHs}). More mathematical details and additional illustrative examples are discussed in the appendices. In this paper, we set the fundamental constants $G = c =1$.

\section{Ladder structure for a general second-order OLDE}\label{Section: Ladder structure for a general second-order differential equation}
In this section, we aim to identify the conditions under which second order OLDEs, which arise in a wide range of physical problems, can be decomposed into two first order pieces forming a ladder structure. Such hierarchical structure, when exists, dramatically simplifies the mathematical description and often exposes deeper physical structures in the concerned system. For this purpose, we consider a general second-order OLDE $H_{\ell}\psi_{\ell}=0$, where the operator $H_{\ell}$ has the following form
\begin{equation} \label{Hl}
H_\ell \equiv -\Delta^2(x)\,\partial_x^2 - \Delta(x)\, p_\ell(x)\,\partial_x + q_\ell(x)\,.
\end{equation}
The discrete index $\ell$ appears in many familiar settings, including the construction of polynomial solutions to differential equations, such as in the case of Legendre polynomials~\cite{arfken2012mathematical}. It can also emerge from the imposition of boundary conditions or constraints, as exemplified by the quantum harmonic oscillator~\cite{Griffiths:1995, Zettili:2009}. Moreover, note that any scaling $H_\ell \rightarrow \Omega_\ell(x)\, H_\ell$ leaves the solution space of the homogeneous equation $H_\ell\, \psi_\ell=0$ invariant. This allows all second-order OLDEs to be expressible with an $\ell$-independent coefficient $\Delta^2(x)$ for the $\partial_x^2$ term.

\textit{(a) Imposition of the ladder structure:} Demanding that $H_\ell$ admits a non-trivial factorization and supports a ladder structure requires the existence of a pair of first-order operators of the general form
\begin{equation}\label{ladder_oparetors}
\begin{split}
        &D_\ell^+ \equiv -\Delta(x)\, f_\ell(x)\, \partial_x + W_\ell^+(x)\,, 
        \\
        &D_\ell^- \equiv \frac{\Delta(x)}{f_{\ell-1}(x)}\, \partial_x + W_\ell^-(x)\,,
        \end{split}
\end{equation}
which act as the raising and lowering operators on the index $\ell$, respectively. As of now $f_{\ell}(x)$ and $W_{\ell}^{\pm}(x)$ are arbitrary functions, which will be fixed later. These operators must further satisfy the ``commutation'' and factorization conditions
\begin{equation}\label{ladder_structure}
\begin{split}
        &H_{\ell+1}\, D_\ell^+ = D_\ell^+\, H_\ell,\quad H_{\ell-1}\, D_\ell^- = D_\ell^-\, H_\ell\,,
        \\
        &H_\ell = D_{\ell-1}^+\, D_\ell^- + E_\ell(x), \quad H_\ell = D_{\ell+1}^-\,D_\ell^+ + \widetilde{E}_\ell(x)\,.
        \end{split}
\end{equation} 
This algebraic structure appears as a generalization of the operator relations from SUSY QM~\cite{Cooper:1994eh}. In fact, \ref{ladder_structure} captures the hierarchical structure between the neighbouring solutions $\{\psi_\ell, \psi_{\ell\pm1}\}$ and factorize $H_\ell$ into two first-order components up to additive functions $\{E_\ell(x),\, \widetilde{E}_\ell(x)\}$ (similar to factorization energies in SUSY QM~\cite{Cooper:1994eh}). We shall also see that these additive functions turn out to be $x$-independent whenever a ladder structure, as given by \ref{ladder_structure}, exists.

\ref{ladder_structure} also suggests that if $H_\ell\, \psi_\ell=0$, then $\psi_{\ell \pm 1} \propto D_\ell^\pm\, \psi_\ell$ must necessarily satisfy $H_{\ell \pm 1}\, \psi_{\ell \pm 1}=0$. This justifies the names raising and lowering operators for $D_\ell^\pm$. However, physical requirements usually limit the spectrum to $\ell \geq \ell_{\rm min}$, preventing the ladder from cascading below the ``ground state'' characterized by the lowest allowed value $\ell_{\rm min}$. In such a case, knowing this ground state alone is sufficient to reconstruct the entire tower of solutions by the repeated application of $D_\ell^+$.

\textit{(b) Existence criterion:} Note that not all second-order OLDEs of the form given by \ref{Hl} generically support such a ladder structure. In fact, the imposition of \ref{ladder_structure} should put restrictions on various functions including $W_\ell^\pm(x)$, which we aim to work out now. We note that \ref{ladder_structure} must hold for arbitrary test functions $\psi_\ell$. Consequently, the resulting coefficients of $\psi_\ell$ and its derivatives, on both sides of each equation, should match separately. Enforcing these identities yields a set of conditions required for the ladder structure to exist. Here, we shall only quote a subset of the relevant final expressions\footnote{Note, we have adopted relations connecting $\{\ell,\ell \pm 1\}$ operators in \ref{ladder_structure}. An entirely analogous analysis can be carried out if one instead considers relations among $\{\ell,\ell \pm n\}$ operators. In this case, we need to simply replace $\ell \pm 1$ by $\ell \pm n$ in all the relations (see \ref{Section: Two examples for $n$-step ladder} for a few examples). \label{f1}} 
\begin{align} \label{conditions}
    &p_{\ell}(x)=p_{\ell-1}(x)-2\, \Delta(x)\,\frac{f'_{\ell-1}(x)}{f_{\ell-1}(x)},\nonumber\\
    &W_\ell^-(x) = -\frac{B_\ell}{f_{\ell-1}(x)}-\frac{\Delta'(x)-p_\ell(x)+J_\ell(x)}{2\, f_{\ell-1}(x)},\nonumber\\
    &W_\ell^+(x) = -B_{\ell+1}\, f_{\ell}(x)-\Delta(x)\, f'_\ell(x)+\frac{f_{\ell}(x)}{2} \times \nonumber\\
    &\kern9.3em \left[\Delta'(x)-p_{\ell+1}(x)-J_{\ell+1}(x)\right],\nonumber\\
    &\widetilde{E}_\ell(x) = E_{\ell+1}(x), \, \, E_\ell(x) = q_\ell(x)-W_\ell^-(x) W_{\ell-1}^+(x), \nonumber\\
    &\kern11em +f_{\ell-1}(x)\, \Delta(x)\, \partial_x W_\ell^-(x),
\end{align}
where $B_\ell$ is so far an arbitrary $x$-independent (integration) constant, the `prime' denotes differentiation with respect to $x$, and $J_\ell(x)$ is given by the following integral:
\begin{equation} \label{Jl}
    J_\ell(x) \equiv \int \frac{\left[H_\ell-q_{\ell-1}(x)\right]\, f_{\ell-1}(x)}{\Delta(x)\, f_{\ell-1}(x)}\, dx.
\end{equation}
Here, the numerator in the integrand should be read as the operator $\left[H_\ell-q_{\ell-1}(x)\right]$ acting on $f_{\ell-1}(x)$. More mathematical details on the above derivation are provided in \ref{appa}. 

A few important comments are in order. If the function $p_\ell(x)$ is specified to be $\ell$-independent, then \ref{conditions} implies a natural choice for $f_\ell(x)$ to be a constant. However, if the function $p_\ell(x)$ is $\ell$-dependent, then $f_\ell(x)$ can be chosen as per \ref{conditions} for a given $\Delta(x)$. We emphasize again that, at the outset, we allowed the functions $\{E_\ell, \widetilde{E}_\ell\}$ to be potentially $x$-dependent and unrelated. However, the imposition of ladder structure has already established a non-trivial relation between them, a ``shape-invariance'' condition closely analogous to SUSY QM~\cite{Cooper:1994eh}. We next show that the final remaining condition, which has not yet been specified, further requires them to be $x$-independent.

It turns out that the imposition of ladder actually gives rise to an overdetermined system (see \ref{appa}), where the number of equations/conditions exceeds the number of unknowns. This, in turn, imposes additional consistency constraints on the functional forms of various quantities appearing in the ladder structure, when it exists. In particular, apart from \ref{conditions}, ladder structure implies two more conditions (check \ref{appa}) on $W_\ell^\pm(x)$. But due to the algebraic relation between $W_\ell^\pm$  imposed by \ref{conditions}, only one of them turns out to be independent. This condition, which we shall refer to as the \textit{``litmus-test criterion''},
\begin{equation} \label{litmus}
    \begin{split}
        &\left[H_\ell-2\, q_\ell(x)+q_{\ell-1}(x)\right]W_\ell^-(x) \\
        &= \frac{\Delta(x)}{f_{\ell-1}(x)}\left[q_\ell'(x)+2\, \Delta(x)\, f_{\ell-1}'(x)\, \partial_xW_\ell^-\right],
    \end{split}
\end{equation}
is in fact the sought-after consistency criterion (as $W_\ell^{-}$ given by \ref{conditions} may not satisfy \ref{litmus}, in general), which connects or constrains various hitherto unrelated functions and parameters. 

In practice, one needs to check whether any values of $\{B_\ell,f_\ell(x)\}$ at all exist that lead to the above condition being met. If no such values can be found, then the underlying system will not admit a ladder structure. In this sense, \ref{litmus} provides a necessary and sufficient condition for the existence of the ladder structure for a second-order OLDE. Interestingly, a straightforward computation of $E'_\ell(x)$ supplemented by the conditions derived in \ref{conditions} and \ref{litmus} further implies $E'_\ell(x)=0$ for all $\ell \geq \ell_{\rm min}$ (hence, $\widetilde{E}'_\ell(x)=E'_{\ell+1}(x)=0$ as well). In other words, the existence of a ladder structure necessarily forces $E_\ell$ to be $x$-independent.

Interestingly, there are various equivalent yet illuminating ways to express \ref{litmus}. For instance, by replacing $W_\ell^-$ from \ref{conditions}, the above litmus-test criterion can be reformulated as \textit{a requirement on $B_\ell$ that makes the following quantity independent of $x$}:
\begin{equation}\label{litmus_test_A}
    \begin{split}
        &- \frac{1}{2}\Delta(x)\, \partial_x\left[\Delta '(x)-p_\ell(x)+J_\ell(x)\right] -B_\ell\, J_\ell(x)\\
        &+q_\ell(x)+\frac{1}{4}\left[\left\{\Delta '(x)-p_\ell(x)\right\}^2 - J_\ell^2(x)\right].
    \end{split}
\end{equation}
This particular version may streamline the process of finding $B_\ell$, which can then be used in \ref{conditions} to obtain $W_{\ell}^\pm(x)$. Another equivalent description of the litmus-test condition is discussed in \ref{Appendix: Generalization of the Darboux condition for second order OLDEs}.

In summary, the construction of the ladder, when it exists, can be achieved as follows. Given any general second-order OLDE, as in \ref{Hl}, first choose $f_\ell(x)$ obeying the first  requirement in \ref{conditions} on $p_\ell(x)$. With this choice, the remaining equations in \ref{conditions} fix the functional forms of $W_\ell^\pm(x)$ and $\{E_\ell(x),\, \widetilde{E}_\ell(x)\}$, up to a so-far-arbitrary $x$-independent constant $B_\ell$. This constant can then be fixed as per the requirement of the litmus test, given by \ref{litmus}. Any inconsistency  encountered in this process (in particular, failing the litmus test) indicates that the given second-order OLDE does not admit a ladder structure in the coordinate $x$ and the solution space spanned by $\psi_\ell(x)$ satisfying $H_\ell\, \psi_\ell=0$.\footnote{This, however, does not necessarily preclude the possibility that after a suitable coordinate transformation $x \to \tilde{x}$ and/or a field redefinition $\psi_\ell(x) \to \widetilde{\psi}_\ell(x)$, the transformed second-order operator $\widetilde{H}_\ell$ may indeed support a ladder structure. In such cases, one needs to repeat the same prescription for $\widetilde{H}_\ell$. Check \ref{appB.2} for a concrete example.}

Our analysis also suggests that the class of OLDEs admitting a ladder structure could be far richer than previously recognized from studies of a limited set of physical examples. Additionally, the existence of ladder structure with respect to the quadruplet $\{I, H_\ell,D_\ell^-,D_{\ell-1}^+\}$, with $I$ being the identity, gives rise to an interesting operator algebra 
\begin{equation} \label{algebra}
    \begin{split}
        &[D_\ell^{-},D_{\ell-1}^+]=\left(H_{\ell-1}-H_{\ell}\right)\, I\equiv Q_\ell\, I,\\
        &[H_{\ell},D_\ell^{-}]=-Q_{\ell}\, D_\ell^-,\, \, [H_{\ell-1},D_{\ell-1}^{+}]=Q_{\ell}\, D_{\ell-1}^+,
    \end{split}
\end{equation}
which follows from \ref{ladder_structure} and the relation $\widetilde{E}_\ell=E_{\ell+1}$. Note that the algebra does not generically close into a Lie algebra. Remarkably, however, the algebra does reduce to a bona fide Lie algebra when $Q_\ell$ becomes $x$-independent (equivalently, when both $f_{\ell-1}$ and $q_\ell -q_{\ell-1}$ are $x$-independent), a special case familiar from standard QM. These observations naturally delineate a subclass of OLDEs that posses this enhanced symmetry and algebraic solvability. Hence, in our view, this work represents a substantive step towards a general understanding and systematic classification of second-order OLDEs based on the existence of ladder symmetry.

\section{Applications}\label{Section: Applications}
We now illustrate the application of our framework on a few important physics problems, namely the quantum harmonic oscillator and dynamical TLNs of Kerr BHs for general spin (both bosonic and fermionic) perturbations. These particular examples are chosen to highlight the broad scope of our method\footnote{See \ref{appB} for additional examples, including (confluent) hypergeometric equations, as well as several corollaries that reproduce known results in literature.}. 

\subsection{Spectrum of quantum harmonic oscillator}\label{Section: Spectrum of quantum harmonic oscillator}
The quantum mechanical description of a simple harmonic oscillator of mass $m$ is governed by the Schrödinger equation~\cite{Griffiths:1995, Zettili:2009},
\begin{equation} \label{Sch}
    -\frac{\hbar^2}{2m} \frac{\mathrm{d}^2\psi_\ell(x)}{\mathrm{d}x^2} + \frac{1}{2} m \omega^2 x^2 \psi_\ell(x) = \mathcal{E}_\ell\, \psi_\ell(x),
\end{equation}
where $\hbar$ is the reduced Planck's constant, $\psi_\ell(x)$ is the wave function, and $\mathcal{E}_\ell$ is the associated energy labelled by the principal quantum number $\ell$. For obvious dimensional reason, we shall rewrite $\mathcal{E}_\ell = \hbar \omega \nu_\ell$. Here, $\nu_\ell$ is some yet-to-be-determined function of $\ell$. A direct comparison with \ref{Hl} provides an identification: $\Delta(x) = \hbar/\sqrt{2m}$, $p_\ell(x)=0$, and $q_\ell(x) = (m\omega^2x^2/2)-\hbar \omega \nu_\ell$. 

Then, from the first equation in \ref{conditions}, we obtain $f_\ell(x)=1$ as the simplest choice, which further fixes $W_\ell^\pm$ and $E_\ell(x)$ up to an $x$-independent parameter $B_\ell$. Now, we need to check the litmus-test criterion given by \ref{litmus} to assess whether the underlying system supports a ladder structure. It boils down to the condition:
\begin{equation*}
    (\nu_\ell-\nu_{\ell-1})\, B_\ell-\sqrt{\frac{m \omega^2}{2}} \left[(\nu_\ell-\nu_{\ell-1})^2-1\right]x=0\,,
\end{equation*}
for all $x \in [-\infty, \infty]$, which can be satisfied if $B_\ell=0$ and  $\nu_\ell=\nu_{\ell-1} \pm 1$. This recurrence relation can be solved as, $\nu_\ell=\pm \ell+c_0$ for some constant $c_0$. However, for spectral stability of the system (i.e., energy should be bounded below), we must choose the positive branch, namely $\nu_\ell= \ell+c_0$. Therefore, \ref{Sch} indeed admits a ladder structure with
\begin{equation} \label{SchDs}
    \begin{split}
        &D_\ell^\pm = \frac{\mp \hbar\, \partial_x+m\omega\, x}{\sqrt{2m}},\quad E_\ell = -\hbar \omega \left(\ell+c_0-\frac{1}{2}\right),
    \end{split}
\end{equation}
which easily follow from \ref{ladder_oparetors} and \ref{conditions}. Note that the above operators match with the standard raising ($D_\ell^+$) and lowering ($D_\ell^-$) operators, up to a constant multiplicative factor~\cite{Griffiths:1995, Zettili:2009}. Also, the ladder triplet $\{H_\ell,D_\ell^-,D_{\ell-1}^+\}$ and the identity operator do form a Lie algebra using \ref{algebra}, since $Q_\ell = \hbar \omega$ is $x$-independent in this case.

Now, to assign a value to $c_0$, we observe that it is related to a shift in the origin of $\ell$. This translational freedom can be fixed by forcing $\ell_{\rm min}=0$ as the origin of the principle quantum number and characterizing the ground state by $H_0 \psi_0 = D_0^- \psi_{0}=0$. Then, the first relation in the second line of \ref{ladder_structure} implies $E_0=0$, thereby fixing $c_0=1/2$. This entails the standard result $\mathcal{E}_\ell=(\ell+1/2)\hbar \omega$ as the energy spectrum of the quantum harmonic oscillator~\cite{Griffiths:1995, Zettili:2009}. The energy eigenstates can also be obtained easily by repeated applications of $D_\ell^+$ operator on the ground state $\psi_0 \sim \exp[-m \omega x^2/(2 \hbar)]$, as required by  $D_0^- \psi_{0}=0$~\cite{Griffiths:1995, Zettili:2009}. 

\subsection{Dynamical tidal response of Kerr BHs}\label{Section: Dynamical tidal response of Kerr BHs}
We now turn to the emergence of ladder structure in the dynamical tidal response of Kerr BHs. The scalar ($s=0$), fermionic ($s=\pm 1/2$), electromagnetic ($s=\pm1$), and gravitational ($s=\pm2$) perturbations of a Kerr BH are governed by a set of decoupled and separable equations, whose radial counterparts are known as the Teukolsky equations~\cite{Teukolsky:1972my,Teukolsky:1973ha,Press:1973zz,Teukolsky:1974yv}. When written in terms of the ingoing Kerr coordinates $\{v,x,\theta,\widetilde{\phi}\}$, the radial part ${}_{s}\psi_{\ell m}(x)$ of the perturbing field
\begin{equation*}\label{TeukField}
    \rho^{-s+|s|} \zeta^{(s)} = \int \frac{\mathrm{d}\omega\,e^{-i\omega v}}{x^s (1-x)^s}\sum_{\ell m} e^{-im\widetilde{\phi}}\,_{s}S_{\ell m}(\theta)\,{}_s\psi_{\ell m}(x),
\end{equation*}
satisfy the following source-free equation in the low-frequency regime $M\omega \ll 1$:
\begin{equation}\label{Teuk2}
    \begin{split}
        &x(1-x)\, {}_s\psi''_{\ell m} + \left[(1-s)(1-2x)+2i P_+ +2i\omega x (2r_+\right.
        \\ 
        &\kern2em \left.-\sigma x)\right] {}_s\psi'_{\ell m}+\left[A_\ell-2i \omega \sigma x (1-2s)\right]{}_s\psi_{\ell m} \approx 0,
    \end{split}
\end{equation}
as we consistently neglect all higher-order terms in $M\omega$. Here, $x=(r_+-r)/\sigma \in (-\infty,0]$ is a dimensionless version of the standard radial coordinate $r \in [r_+,\infty)$ with $\sigma=(r_+ -r_-)$ (assuming a non-extremal BH), $\omega \in \mathbb{R}$ is the perturbation frequency, $\rho = -(r-i a \cos\theta)$, $\,_{s}S_{\ell m}(\theta)$ is the spin-weighted spheroidal harmonics, and $r_\pm= M\pm \sqrt{M^2-a^2}$ denote the positions of the event and Cauchy horizons of the Kerr BH with mass $M$ and spin $a$. Moreover, we have $2P_+ = -(\widetilde{\omega}_+/\kappa_+)$ with $\widetilde{\omega}_+ = (\omega-m \Omega_+)$, where $\Omega_+ = a/(2Mr_+)$ and $\kappa_+=\sigma/(2Mr_+)$ are the angular speed and surface gravity of the event horizon, respectively. The quantity $A_\ell=\lambda + 2s +2i\omega r_+(1-2s)$ is the redefined separation constant, with $\lambda$ till $\mathcal{O}(M\omega)$ is given by~\cite{Teukolsky:1972my,Teukolsky:1973ha,Press:1973zz,Teukolsky:1974yv}
\begin{equation} \label{lambda}
        \lambda \approx (\ell-s) (\ell+s+1)-2am\omega \left\{1+\frac{s^2}{\ell(\ell+1)}\right\}.
\end{equation} 
To proceed, we seek the near-horizon ($0<x \ll 1$) solution obeying ingoing boundary condition. To this end, we first neglect terms proportional to $M\omega x^n$ in \ref{Teuk2} for all $n \in \mathbb{N}$ and then\footnote{The ordering of these two operations are absolutely crucial and should be done as described in the main text. Otherwise one will artificially neglect some terms in the perturbation equation.} multiply the resulting equation by an overall factor of $x(x-1)$ for easy comparison with the canonical form of \ref{Hl}. It leads to the identifications: $\Delta(x) = x(1-x)$, $p_\ell(x) = (1-s)(1-2x)+2i P_+$, and $q_\ell(x) = -x(1-x)\, A_\ell$. Then, the litmus-test condition in \ref{litmus} can be satisfied if we choose $f_\ell(x)=1$ and $B_\ell = (\ell^2-2is P_+)/(2\ell)$. This, in turn, suggests that a ladder structure exists with
\begin{equation} \label{BHE}
    E_\ell \approx -\frac{(\ell^2-s^2)(\ell^2+4P_+^2)}{4\ell^2},
\end{equation}
where the value of $P_+^2$ should be taken till $\mathcal{O}(M\omega)$. One can also compute other relevant quantities, like $D_\ell^{\pm}$, following \ref{conditions}:
\begin{equation} \label{BHD}
    \begin{split}
        &D_\ell^+ \approx x(x-1) \partial_x - \frac{(\ell-s+1)[(\ell+1)(1-2x)+2iP_+]}{2(\ell+1)},\\
        &D_\ell^- \approx x(1-x) \partial_x - \frac{(\ell+s)[\ell(1-2x)-2iP_+]}{2\ell}.
    \end{split}
\end{equation}
At this stage, we must mention a cautionary note regarding our calculation above. The litmus-test condition yields an equation of the schematic form $(B_\ell+\cdots)x+\mathcal{O}(M\omega x^2, M^2\omega^2)=0$. This renders the identification of the $\mathcal{O}(M\omega)$-part of $B_\ell$ ambiguous, since any such contribution produces a term proportional to $M\omega x$ in the litmus-test equation that can be neglected under our perturbation scheme. However, for the sake of concise presentation, we have selected the aforementioned choice of $B_\ell$ that leads to a compact expression for $E_\ell$. In fact, as we shall see below that the response function ${}_sF_{\ell m}(\omega)$ in \ref{Fl} has a multiplicative ambiguity ${}_s\alpha_{\ell m}(\omega)$, different choices of $B_\ell$ that differ by $\mathcal{O}(M\omega)$-contributions therefore correspond to different normalizations ${}_s\alpha_{\ell m}(\omega)$ when calibrated with, say, the corresponding effective field theory (EFT)/BH perturbation theory (BHPT) results~\cite{Porto:2016pyg, Porto:2016zng, Charalambous:2021mea, Ivanov:2022hlo, Saketh:2023bul, Bhatt:2023zsy, Bhatt:2024rpx, Chakraborty:2025wvs, Combaluzier--Szteinsznaider:2025eoc}. In this sense, our earlier choice of $B_\ell$ is entirely equivalent to any other within this freedom. However, for the sake of completeness, we will later discuss the case with a different choice (free of $M\omega$ term) of $B_\ell$ satisfying the litmus-test condition.

With the above ladder operators in hand, we now turn to the computation of the tidal response of the BH. We start by constructing the response function ${}_sF_{|s|m}$ for the ground state ($\ell=|s|$) obeying $\partial_x\,{}_sF_{|s| m}=0$. From the perturbation equation $H_{|s|}\, {}_s\psi_{|s| m} \equiv \partial_x\, {}_sF_{|s|m} =0$, we can identify 
\begin{equation} \label{Fs}
    \begin{split}
        {}_sF_{|s|m} &= x(1-x)\, {}_s\psi'_{|s| m}+(2iP_+-s)\, {}_s\psi_{|s| m} \\
        &\kern1em +2s \int \textrm{d}x\, x\, {}_s\psi'_{|s| m} + A_{|s|} \int \textrm{d}x\, {}_s\psi_{|s| m}~.
    \end{split}
\end{equation}
Here, ${}_s\psi_{|s| m}$ is chosen to satisfy the ingoing boundary condition at the horizon. Then, the above expression simplifies to ${}_sF_{|s|m} \sim (2iP_+-s)$, up to some multiplicative/additive ambiguity. However, the form can be further constrained by demanding concordance with the static ($\omega \to 0$) case, for which the response function must reduce to its known purely imaginary (real) form for bosonic (fermionic) perturbations (see Ref.~\cite{Chakraborty:2025zyb} and references therein). This behavior is recovered by choosing ${}_sF_{|s|m} \propto 2i^{1+2s}P_+$. In addition, due to Kerr symmetries (axisymmetry and time-reversal), we must have ${}_sF^*_{\ell m}(\omega)={}_sF_{\ell -m}(-\omega)$~\cite{Futterman:1988ni}. However, the choice $2i^{1+2s}P_+$ does not satisfy this identity for the fermionic case. Motivated by the analysis of Ref.~\cite{Chakraborty:2025zyb}, a simple way to restore the symmetry is to instead consider ${}_sF_{|s|m} \propto 2i^{1+2s} \left(P_+\right)^\mu$, where $\mu = 1\, (0)$ for bosonic (fermionic) perturbations. This modified form is also consistent with the absence of superradiance for fermionic fields~\cite{Brito_2020}.

The above construction can be easily generalized to construct the response function ${}_sF_{\ell m}$ of the $\ell$-th ($\ell \geq |s|$) state such that $\partial_x\, {}_sF_{\ell m}=0$. For this purpose, we simply replace ${}_s\psi_{|s| m}$ in \ref{Fs} by $D_{|s|+1}^{-} D_{|s|+2}^- \cdots D_{\ell}^- {}_s\psi_{\ell m}$, where ${}_s\psi_{\ell m}$ is the $\ell$-th state obeying $H_\ell\, {}_s\psi_{\ell m} =0$ and ingoing boundary condition at the horizon. In fact, we can compute it by repeated application of the raising operator given by \ref{BHD} on the ground state itself, i.e., ${}_s\psi_{\ell m} \propto D_{\ell-1}^{+} D_{\ell-2}^+ \cdots D_{|s|}^+\, {}_s\psi_{|s| m}$. Then, using the last equation in \ref{ladder_structure} and \ref{BHE}, we obtain
\begin{equation} \label{Fl}
    \begin{split}
        &{}_sF_{\ell m}(\omega)=2i^{1+2s}  \left(P_+\right)^\mu\, {}_s\alpha_{\ell m} (\omega) \prod_{k=|s|+1}^{\ell} E_k\\
        &=\frac{2 \left(P_+\right)^\mu\,{}_s\alpha_{\ell m} (\omega)}{i^{-(1+2s)}4^{\ell-|s|}} \frac{\displaystyle \binom{\ell+|s|}{|s|}}{\displaystyle \binom{\ell}{|s|}\, \binom{2|s|} {|s|}} \frac{|\Gamma(\ell+2iP_++1)|^2}{|\Gamma(|s|+2iP_++1)|^2},
    \end{split}
\end{equation}
where the right hand side of the above expression should be taken till $\mathcal{O}(M\omega)$ and ${}_s\alpha_{\ell m}$ is a constant, whose functional form cannot be fixed within the ambit of the ladder computation itself and should be calibrated instead using inputs from effective field theory (EFT) and BH perturbation theory~\cite{Porto:2016pyg, Porto:2016zng, Charalambous:2021mea, Ivanov:2022hlo, Saketh:2023bul, Bhatt:2023zsy, Bhatt:2024rpx, Chakraborty:2025wvs, Combaluzier--Szteinsznaider:2025eoc}. Nevertheless, imposition of ${}_sF^*_{\ell m}(\omega)={}_sF_{\ell -m}(-\omega)$ implies ${}_s\alpha^*_{\ell m}(\omega)= {}_s\alpha_{\ell -m}(-\omega)$. 

Now, if ${}_s\alpha_{\ell m}$ does not depend on BH/perturbation parameters (i.e., it depends only on $\ell, m$, and $s$), its value must coincide with that in the Schwarzschild limit, where spherical symmetry enforces $m$-independence. This fixes ${}_s\alpha_{\ell m} \equiv {}_s\alpha_{\ell}$ to be independent of $m$. Combined with the condition above, we conclude that ${}_s\alpha_{\ell}$ must be real for both bosonic and fermionic cases. Then, the tidal response function ${}_sF_{\ell m}$ becomes purely imaginary (real) in the bosonic (fermionic) case. As a result, the bosonic TLNs for Kerr BHs will vanish till $\mathcal{O}(M\omega)$. However, an increasing body of works~\cite{Porto:2016pyg, Porto:2016zng, Charalambous:2021mea, Ivanov:2022hlo, Saketh:2023bul, Bhatt:2023zsy, Bhatt:2024rpx, Chakraborty:2025wvs, Combaluzier--Szteinsznaider:2025eoc} have already suggested that the dynamical bosonic TLNs of Kerr BH do not generically vanish (e.g., for non-axisymmetric perturbations). This, in turn, suggests that ${}_s\alpha_{\ell m}$ will indeed depend on $(a,\omega)$ for the bosonic case. In general, its explicit form can be determined by matching \ref{Fl} to the corresponding EFT/BHPT result~\cite{Porto:2016pyg, Porto:2016zng, Charalambous:2021mea, Ivanov:2022hlo, Saketh:2023bul, Bhatt:2023zsy, Bhatt:2024rpx, Chakraborty:2025wvs, Combaluzier--Szteinsznaider:2025eoc}.

To the best of our knowledge, this is the first consistent derivation of the dynamical spin-$s$ tidal response function of a Kerr BH based on ladder symmetry, with all intermediate subtleties carefully treated. Our approach differs sharply from earlier analyses in two important respects. First, unlike the usual Boyer–Lindquist frame, our computation in the ingoing Kerr coordinates yields a smooth, horizon-penetrating description of the near-horizon region, providing a corotating frame in which a Newtonian-like interpretation of stationary tides remains meaningful. Second, we work directly with the Teukolsky equations governing Weyl curvature perturbations, instead of metric perturbations. This ensures a fully gauge-invariant characterization of the tidal response for both bosonic ($s = 0,~\pm1,~\pm2)$ and fermionic ($s = \pm 1/2)$ perturbations. In fact, the multiplicative ambiguity ($\,_s\alpha_{\ell m}$) of the tidal response function in \ref{Fl} can be properly chosen to match with the corresponding results obtained via other approaches. More broadly, our derivation reveals that the presence of ladder symmetry is not limited to the class of 2nd order OLDEs traditionally associated with special functions such as the (confluent) hypergeometric family.

We now return to the point related to the ambiguity of the $M\omega$-part in $B_\ell$ and discuss the result for a different choice of $B_\ell$. As discussed earlier, the schematic form of the litmus-test condition is $(B_\ell+\cdots)x+\mathcal{O}(M\omega x^2, M^2\omega^2)=0$. Thus, under our approximation scheme, another well-motivated solution is $B_\ell^{\text{(new)}} = (\ell^2-2isP_+^{(0)})/(2\ell)$, which is $M\omega$-independent with $P_+^{(0)} = P_+(\omega=0)$. Following the procedure described previously in this section, the response function now takes the form up to $\mathcal{O}(M\omega)$:
\begin{equation}\label{conserved_qunatity_BH}
    \begin{split}
        &_sF^{\text{(new)}}_{\ell m}(\omega) \approx  \frac{2 i^{1+2s}\, _s\alpha^{\text{(new)}}_{\ell m}(\omega)}{4^{\ell-|s|}} \frac{\displaystyle \binom{\ell+|s|}{|s|}}{\displaystyle \binom{\ell}{|s|}\, \binom{2|s|} {|s|}} \times\\
        &\frac{|\Gamma(\ell+2iP_+^{(0)}+1)|^2}{|\Gamma(|s|+2iP_+^{(0)}+1)|^2}\Bigg[\left(P_+\right)^\mu +\frac{4i\omega \left(2iP_+^{(0)}-s\right)}{\kappa_+}\, _s\Sigma_{\mu \ell}\Bigg].
    \end{split}
\end{equation}
Here, $\,_s\alpha^{\text{(new)}}_{\ell m}(\omega)$ is an arbitrary multiplication factor and $_s\Sigma_{\mu \ell} = \{P_+^{(0)}\}^\mu \sum_{k=|s|+1}^{\ell}{k^2}/[{(k^2-s^2)\{k^2+4 (P_+^{(0)})^2\}}]$. Note that the above expression looks starkly different from that in \ref{Fl}. In fact, unlike the purely imaginary nature of $(_sF_{\ell m}/_s\alpha_{\ell m})$, the new ratio $(_sF^{\text{(new)}}_{\ell m}/_s\alpha^{\text{(new)}}_{\ell m})$ is complex valued. However, these seemingly different results are actually consistent with each other due to the freedom in choosing the parameters $\,_s\alpha_{\ell m}$ and $_s\alpha^{\text{(new)}}_{\ell m}$. Ultimately, these response functions will reduce to a unique value once properly calibrated with the EFT/BHPT result~\cite{Porto:2016pyg, Porto:2016zng, Charalambous:2021mea, Ivanov:2022hlo, Saketh:2023bul, Bhatt:2023zsy, Bhatt:2024rpx, Chakraborty:2025wvs, Combaluzier--Szteinsznaider:2025eoc}. 

\section{Discussion}\label{Section: Discussion}
In the BHPT literature, there is a conception that the vanishing of TLNs is intimately tied to the existence of ladder symmetries, by which the induced multipole moments (whose influence generally decay with distance) of BHs are set to zero. However, this work clearly demonstrates, among other results, that this conception is not unconditionally true. In particular, for dynamical tidal perturbations of Kerr BHs, we show that a ladder structure does exist till the linear order in $M\omega$, even though the corresponding TLN is non-vanishing for non-axisymmetric perturbations. Moreover, the ladder structure alone is unable to fix the values TLNs, owing to inherent multiplicative/additive ambiguities. A matching with the corresponding EFT/BHPT results is therefore essential. Thus, the existence of a ladder symmetry by itself neither entails vanishing TLNs, nor does it provide their actual values. 

The above implication follows as a consequence of more broad results obtained in this work. We have analyzed the existence of ladder structures for the most general second-order OLDEs and derive the following conclusions:

\begin{itemize}

\item Any second-order OLDE supports a ladder structure if it satisfies the ``litmus-test criterion'' as outlined in and around \ref{litmus} and \ref{litmus_test_A}. It provides a practical diagnostic tool that can immediately ascertain whether a given OLDE, describing one's physical system of interest, admits a ladder structure. For example, using this criterion, we have shown in \ref{appB} that both hypergeometric and confluent hypergeometric differential equations admit ladder structures under certain conditions on their parameters. Since these types of equations appear frequently in BH physics, this naturally explains the emergence of ladder structures in the context of static BH perturbations \cite{Hui:2020xxx}.

\item We have illustrated our formalism through applications in quantum physics (the simple harmonic oscillator) and in gravitational physics (the dynamical tidal response of a Kerr BH). In the latter case, we have established, contrary to common belief, that the presence of a ladder structure in BH perturbation equations does not, by itself, imply the vanishing of TLNs.

\item We have also pointed out that the Noether-like conserved charge $_sF_{\ell m}$ associated with Kerr ladder symmetry is not fixed completely, but has a freedom in terms of the multiplicative factor $_s\alpha_{\ell m}$. Thus, we need matching with some other method like EFT/BHPT or, additional physics inputs to completely fix $_sF_{\ell m}$.
\end{itemize}

Before concluding, we would like to emphasize the broad scope of applicability of our work. Since most physical systems are governed by second-order OLDEs (including BH perturbation theory), our framework applies widely across such systems. As demonstrated, it encompasses both quantum mechanical systems, such as the harmonic oscillator, and BH perturbations resulting in their dynamical tidal deformability. Our work also sheds light on the underlying symmetry algebra, and its clear correspondence with supersymmetric theories. For the future, there are ample scopes of extending our framework to further incorporate other physical aspects, including those with inhomogeneous equations having source terms and non-isolated systems like BHs embedded in dark-matter halos. Also, is it possible to generalize our dynamical Kerr ladders order-by-order in $M\omega$ expansion? And, perhaps more importantly, what are the implications of ladder symmetries for the Mano-Suzuki-Takasugi formalism and BH scattering amplitudes? We aim to produce follow ups along these important directions in near future.  

\begin{acknowledgements}
R.G. wants to thank Valerio De Luca for useful discussions. R.G. is supported by the Fulbright-Nehru Postdoctoral Research Fellowship (Award No.3174/FNPDR/2025) from the United States-India Educational Foundation. R.G. also extends his gratitude to the Astrophysical Relativity group at ICTS and acknowledges support of the Department of Atomic Energy, Government of India, under project nos. RTI4019 and RTI4013 during his postdoctoral tenure at ICTS. The research of S.C. is supported by MATRICS (MTR/2023/000049) and Core Research Grants (CRG/2023/000934) from SERB, ANRF, Government of India. SC also thanks the local hospitality at ICTS and IUCAA through the associateship program, where a part of this work was done. R.P.B. and S.B. thank IACS Kolkata for the hospitality during their visits, where a part of this work was done. S.B. acknowledges support from NSF Grant PHY-2309352.
\end{acknowledgements}

\onecolumngrid
\vspace{0.25in}
\rule{\textwidth}{1pt}
\appendix
\labelformat{section}{Appendix #1} 
\labelformat{subsection}{Appendix \thesection.#1}

\section{Derivation of the ladder structure} \label{appa} 
In this Appendix, we provide more details for the derivation of \ref{conditions} and \ref{litmus}. For this purpose, let us start from the operator relations provided in \ref{ladder_structure}. As discussed in the main text, we equate all coefficients of some test function $\phi_\ell$ and its derivatives on both sides of these relations giving rise to the following set of equations. Factorization of $H_\ell$ in terms of $D_{\ell-1}^+$ and $D_\ell^-$, i.e., the first relation in the second line of \ref{ladder_structure}, implies two equations:
\begin{equation} \label{fact1}
    \begin{split}
        &W^+_{\ell-1}(x)= f_{\ell-1}^2(x)\left[W_\ell^-(x)+\partial_x\left\{\frac{\Delta(x)}{f_{\ell-1}(x)}\right\}\right]-f_{\ell-1}(x)\, p_\ell(x),\\
        &q_\ell(x)=W^-_\ell(x)W^+_{\ell-1}(x)-f_{\ell-1}(x)\, \Delta(x)\, \partial_x W^-_\ell(x)+E_\ell(x).
    \end{split}
\end{equation}
The last relation in \ref{conditions} is just a rewriting of the second equation above. On the other hand, the factorization of $H_\ell$ in terms of $D_{\ell+1}^-$ and $D_\ell^+$, i.e., the second relation in the second line of \ref{ladder_structure}, implies (by relabelling $\ell \to \ell-1$)
\begin{equation} \label{fact2}
    \begin{split}
        &W^+_{\ell-1}(x)= f_{\ell-1}^2(x)\left[W_\ell^-(x)+\frac{\partial_x\left\{\Delta(x)\, f_{\ell-1}(x)\right\}}{f^2_{\ell-1}(x)}\right]-f_{\ell-1}(x)\, p_{\ell-1}(x),\\
        &q_{\ell-1}(x)=W^-_\ell(x)W^+_{\ell-1}(x)+\frac{\Delta(x)}{f_{\ell-1}(x)}\, \partial_x W^+_{\ell-1}(x)+\widetilde{E}_{\ell-1}(x).
    \end{split}
\end{equation}
The consistency between $W_{\ell-1}^+$'s obtained from \ref{fact1} and \ref{fact2} above, yields the recurrence relation in $p_\ell(x)$ mentioned in \ref{conditions}. This recurrence can indeed be solved as follows for a unit-step ladder
\begin{equation}\label{fact3}
    p_\ell(x) = p_{\ell_{\rm min}}(x)-2\, \Delta(x) \sum_{k=\ell_{\rm min}}^{\ell-1} \frac{f'_k(x)}{f_k(x)}.
\end{equation}
For an $n$-step ladder, we need to replace $(\ell \pm 1) \to (\ell \pm n)$ in all expressions and the domain of the sum in the right hand side of \ref{fact3} needs to be modified to $k \in \{\ell_{\text{min}}, \ell_{\text{min}}+n, \ell_{\text{min}}+2n, \cdots, \ell_{\text{min}}+(j-1)n\}$, where $\ell = \ell_{\text{min}}+j\, n$.

Now, to obtain the second and third equations in \ref{conditions}, we need to integrate the last equations in \ref{fact1} and \ref{fact2}. In decoupled form, these are a pair of Riccati equations, as also appear in the case of SUSY QM~\cite{Cooper:1994eh}. They can be transformed into two simplified equations, namely
\begin{equation} \label{fact5}
    \begin{split}
        H_{\ell-1}\, \widetilde{W}^+_{\ell-1}(x) = E_{\ell}(x)\, \widetilde{W}^+_{\ell-1}(x),\quad H_\ell\, \widetilde{W}^-_\ell(x) = E_\ell(x)\, \widetilde{W}^-_\ell(x),
    \end{split}
\end{equation}
by the applications of Cole-Hopf substitutions~\cite{Hopf:1950, Cole:1951}
\begin{equation} \label{fact6}
    \begin{split}
        W^+_{\ell-1}(x)=f_{\ell-1}(x)\, \Delta(x)\frac{\partial_x \widetilde{W}^+_{\ell-1}(x)}{\widetilde{W}^+_{\ell-1}(x)},\quad W^-_\ell(x)=-\frac{\Delta(x)}{f_{\ell-1}(x)}\frac{\partial_x \widetilde{W}^-_\ell(x)}{\widetilde{W}^-_\ell(x)}.
    \end{split}
\end{equation}
In terms of these tilde-variables, the first equation in \ref{fact1} can then be integrated to obtain
\begin{equation} \label{fact7}
    \widetilde{W}^+_{\ell-1}(x)\, \widetilde{W}^-_{\ell}(x) = A_\ell\, \frac{\Delta(x)}{f_{\ell-1}(x)}\, \textrm{Exp}\left[-\int \frac{p_\ell(x)}{\Delta(x)}\, dx\right],
\end{equation}
where $A_\ell$ is an $x$-independent integration constant. Now, the aforementioned Riccati equation for $\widetilde{W}_\ell^-(x)$ can be reduced to a first order differential equation as
\begin{equation}\label{Eq_req_A}
    \left[H_\ell - q_{\ell-1}(x)\right]f_{\ell-1}(x) = -\Delta(x)f_{\ell-1}(x)\left[\Delta ''(x)- p_{\ell}'(x)+2\partial_x\left\{f_{\ell-1}(x)W_\ell^-(x)\right\}\right],
\end{equation}
which can now be solved easily by integration (with a constant of integration $B_\ell$) as in \ref{conditions}. Since $W_{\ell-1}^+(x)$ and $W_\ell^-(x)$ are related to each other via \ref{fact1}, we can use it to calculate $W_\ell^+(x)$. Next, the ``commutation'' relations between $H_\ell$ and $D_\ell^\pm$ lead to three new equations involving $W_\ell^\pm$ and $f_\ell$, which makes the whole system overdetermined:
\begin{equation} \label{fact4}
    \begin{split}
        &\left[H_\ell-2\, q_\ell(x)+q_{\ell-1}(x)\right]W_\ell^-(x) = \frac{\Delta(x)}{f_{\ell-1}(x)}\left[q_\ell'(x)+2\, \Delta(x)\, f_{\ell-1}'(x)\, \partial_x W_\ell^-\right],\\
        &\left[H_{\ell-1}-2\, q_{\ell-1}(x)+ q_{\ell}(x)\right]W_{\ell-1}^+(x) = -2\, \Delta^2(x)\, \frac{f_{\ell-1}'(x)}{f_{\ell-1}(x)}\, \partial_x W_{\ell-1}^+(x)-\Delta(x)\, f_{\ell-1}(x)\, q_{\ell-1}'(x),\\
        &\left[H_{\ell-1}-2\, q_{\ell-1}(x)+ q_{\ell}(x)\right]f_{\ell-1}(x) = -2\, \Delta^2(x)\, \frac{f_{\ell-1}'^2(x)}{f_{\ell-1}(x)}+\Delta(x)\, f_{\ell-1}(x)\left[\Delta''(x)-p_{\ell-1}'(x)-2\, \partial_x\left\{\frac{W_{\ell-1}^+(x)}{f_{\ell-1}(x)}\right\}\right].
    \end{split}
\end{equation}
The second and third equations, when supplemented with \ref{fact1} and \ref{fact2}, both imply $\widetilde{E}_{\ell}=E_{\ell+1}$. Whereas the remaining equation is the litmus-test criterion given by \ref{litmus}. This completes our derivation of the central relations used in the main text. We note that one could have proceeded a bit differently (e.g., starting by manipulating the ``commutation'' relations first). However, at the end, one indeed gets the similar set of conditions as mentioned above.

\section{More examples of the ladder structure}\label{appB}
In this appendix, we will discuss more examples and some corollaries of our framework presented in \ref{Section: Ladder structure for a general second-order differential equation}.

\subsection{Two examples for $n$-step ladder}\label{Section: Two examples for $n$-step ladder}
In this subsection, we demonstrate two examples, namely the static scalar field perturbation equations of Schwarzschild-Tangherlini (ST) and analog (AN) BH, where the underlying 2nd order OLDEs support  ladder structures that connects $\ell \to \ell \pm n$ (we shall see that $n = \pm 1$ does not work, in contrast to earlier examples) states in the solution space of $H_\ell \psi_\ell=0$. Though both of these examples are known in literature, see for example Refs.~\cite{Berens:2025jfs, DeLuca:2025zqr}, we consider them as illustrative examples in our framework. The corresponding Hamiltonian operators in these two cases can be identified with \ref{Hl} for the choices
\begin{equation}
    \begin{split}
        &\Delta^{(ST)}(x) = x(x^{D-3}-x_+^{D-3}), \quad p^{(ST)}_\ell(x)= \partial_x \Delta^{(ST)}(x),\quad q^{(ST)}_\ell(x) = \ell(\ell+D-3)\, x^{D-4}\, \Delta^{(ST)}(x), \\
        &\Delta^{(AN)}(x) = x(x^{4}-x_+^{4}), \quad p^{(AN)}_\ell(x)= \partial_x \Delta^{(AN)}(x)-\frac{3 \Delta^{(AN)}(x)}{x},\quad q^{(AN)}_\ell(x) = \ell(\ell+1)\, x^{3}\, \Delta^{(AN)}(x),
    \end{split}
\end{equation}
where $D \geq 4$ represents the spacetime dimension in the ST case, and $x_+$ represents the locations of the event horizons, which we represent by the same parameter $x_+$ in both cases by abuse of notations. Also, since $p_\ell(x)$ is independent of $\ell$ in both cases, we set $f_\ell(x)=1$. Then, a straightforward application of the litmus-test condition in \ref{litmus} for the existence of $n$-step ladder (see Footnote-\ref{f1}) implies
\begin{equation}
    \begin{split}
        &B_\ell^{(ST)} = -\frac{(D-3) \left[2\ell(\ell+D-3)-n(2\ell+D-3)+n^2\right] x_+^{D-3}}{2n(2\ell-n+D-3)},\quad n \overset{\text{(ST)}}{=} \pm (D-3),\\
        &B_\ell^{(AN)} = -\frac{2 \left[2\ell(\ell+1)-n(2\ell+1)+n^2-3\right] x_+^{4}}{n(2\ell-n+1)},\quad n \overset{\text{(AN)}}{=} \pm 4,
    \end{split}
\end{equation}
Other associated quantities can be computed too using \ref{conditions}, which we shall skip here.

\subsection{Generalization of the Darboux condition for 2nd order OLDEs admitting ladder symmetry}\label{Appendix: Generalization of the Darboux condition for second order OLDEs}
In this appendix, we aim to rewrite our litmus-test criterion in yet another suggestive form that helps us identify a Darboux-like condition for the existence of an $n$-step ladder. For this purpose, we first express \ref{litmus_test_A} [or equivalently \ref{litmus}] as $2B_\ell = \left[S_\ell'(x)/J_\ell'(x)\right] -J_\ell(x)$ with
\begin{equation}
    S_\ell(x) \equiv \left[q_\ell(x)+q_{\ell-n}(x)\right]-\underbrace{\frac{[H_\ell-q_\ell(x)]f_{\ell-n}(x)}{f_{\ell-n}(x)}}_{\equiv \mathcal{G}_\ell(x)}+\frac{1}{2}\left[p_\ell^2(x)-\Delta'^2(x)\right]+\Delta^2(x)\, \partial_x \left[\frac{p_\ell(x)-\Delta'(x)}{\Delta(x)}\right],
\end{equation}
and $J_\ell(x)$ is given by \ref{Jl} with the index $(\ell-1) \to (\ell-n)$. Here, our notation $[H_\ell(x)-q_\ell(x)] X(x)$ stands for the operator $[H_\ell(x)-q_\ell(x)]$ acting on $X(x)$, which can be expressed more explicitly as $-\Delta^2(x)\, X''(x) - \Delta(x)\, p_\ell(x)\, X'(x)$. Now, as discussed before, $B_\ell$ needs to be $x$-independent (i.e., $\partial_x B_\ell = 0$) for the existence of ladder. After some algebraic manipulations, this condition leads us to
\begin{equation}\label{generalised_Darboux}
\partial_x\left[\frac{\delta q_{\ell}'(x)}{\widetilde{\delta{q}}_\ell(x)}\right] + 2\, \partial_x \left[\frac{q_{\ell-n}'(x)}{\widetilde{\delta{q}}_\ell(x)}\right] -\widetilde{\delta{q}}_\ell(x)
-\partial_x\left[\frac{[H_\ell(x)-q_\ell(x)]\left\{p_{\ell}(x)-\Delta '(x)\right\}}{\Delta(x)\widetilde{\delta{q}}_\ell(x)}\right]-\partial_x\left[\frac{\mathcal{G}'_\ell(x)}{\widetilde{\delta{q}}_\ell(x)}\right] = 0,
\end{equation}
where $\delta{q_\ell(x)} \equiv q_\ell(x)-q_{\ell-n}(x)$ and
\begin{equation}
    \Delta(x)\, \widetilde{\delta{q}}_\ell(x) \equiv \frac{[H_\ell(x)-q_{\ell-n}(x)]f_{\ell-n}(x)}{f_{\ell-n}(x)}=\delta q_\ell(x)+\mathcal{G}_\ell(x).
\end{equation}

As a concrete application, let us apply \ref{generalised_Darboux} on one of the most ubiquitous differential equations in physics, namely the Schr\"odinger-type equation with Hamiltonian $H_\ell = -\partial_x^2 + V_\ell(x)$. Here, $V_\ell(x)$ denotes the potential. Now, we derive the condition on $V_\ell(x)$ under which the system admits an $n$-step ladder. Comparing the Schr\"odinger-like Hamiltonian with \ref{Hl}, we obtain $\Delta(x) = 1,~p_\ell(x) = 0,~q_\ell(x) = V_\ell(x)$. Then, assuming $f_\ell(x) = {\rm constant}$, \ref{generalised_Darboux} reduces to
\begin{equation}
\partial_x\left[\frac{\delta{V_\ell'(x)}}{\delta{V_\ell(x)}}\right] + 2\, \partial_x \left[\frac{V_{\ell-n}'(x)}{\delta{V_\ell(x)}}\right] -\delta{V_\ell(x)} = 0,
\end{equation}
where $\delta{V_\ell(x)} \equiv V_\ell(x)-V_{\ell-n}(x)$. This expression matches exactly with \textit{the Darboux condition} derived in Ref.~\cite{DeLuca:2025zqr} for the existence of an $n$-step ladder in a Schr\"odinger-type system. In the same spirit, \ref{generalised_Darboux} can be considered as {\em the generalization of this Darboux condition for any 2nd order OLDE admitting a ladder symmetry.}

\subsection{Ladder in hypergeometric differential equation}\label{appB.1}
The hypergeometric equation has the form:
\begin{equation} \label{hyp}
    x(1-x)\, \psi_\ell''(x) + [c_\ell-(a_\ell+b_\ell+1)x]\, \psi_\ell'(x)-a_\ell b_\ell\, \psi_\ell(x)=0.
\end{equation}
After multiplying by $x(x-1)$, it can be written as \ref{Hl}, with
\begin{equation}\label{Hypergeometric_Hamiltonian_comarison}
    \Delta(x) = x(x-1),\qquad p_{\ell}(x) = (a_\ell+b_\ell+1)x - c_\ell, \qquad q_{\ell}(x) = -a_\ell b_\ell\, x(x-1).
\end{equation}
If we consider $f_\ell(x) = 1$, then the first equation in \ref{conditions} suggests both $c_\ell$ and $(a_\ell+b_\ell)$ should be $\ell$-independent, and the litmus-test condition in \ref{litmus} for the existence of ladder further implies that
\begin{equation}
    a_{\ell+1} -a_{\ell} = \pm 1, \qquad {\rm or}\qquad  b_{\ell+1} - a_{\ell} = \pm 1.
\end{equation}
Then, the relevant quantities can be straightforwardly worked out using \ref{conditions}. These expressions are summarized in \ref{table_Hypgeo_Ladder_2}.  

\begin{table}[H]
    \centering
    \setlength{\tabcolsep}{20pt}
    \renewcommand{\arraystretch}{2}
    \begin{tabular}{cccc}
    \hline
         Cases & $W_{\ell}^+(x)$ & $W_{\ell}^-(x)$ & $E_{\ell}(x)$ \\ \hline
         $a_{\ell+1} = a_{\ell} + 1$ &  $-a_{\ell}x-\frac{a_{\ell} (b_{\ell}-c_h)}{a_{\ell}-b_{\ell}+1}$ & $b_{\ell}x-\frac{b_{\ell} (a_{\ell}-c_h)}{a_{\ell}-b_{\ell}-1}$ & $- \frac{(a_{\ell}-1) b_{\ell} (c_h-a_{\ell}) (-b_{\ell}+c_h-1)}{(-a_{\ell}+b_{\ell}+1)^2}$\\
        
        $a_{\ell+1} = a_{\ell} - 1$ &  $-b_{\ell}x+\frac{b_{\ell} (a_{\ell}-c_h)}{a_{\ell}-b_{\ell}-1}$ & $a_{\ell}x+\frac{a_{\ell} (b_{\ell}-c_h)}{a_{\ell}-b_{\ell}+1}$ & $- \frac{a_{\ell} (b_{\ell}-1) (-a_{\ell}+c_h-1) (c_h-b_{\ell})}{(a_{\ell}-b_{\ell}+1)^2}$\\ 
        
        $a_{\ell+1} = b_{\ell} + 1$ & $-b_{\ell}x+\frac{b_{\ell} (a_{\ell}-c_h)}{a_{\ell}-b_{\ell}-1}$ & $b_{\ell}x-\frac{b_{\ell} (a_{\ell}-c_h)}{a_{\ell}-b_{\ell}-1}$ & $- \frac{(a_{\ell}-1) b_{\ell} (c_h-a_{\ell}) (-b_{\ell}+c_h-1)}{(-a_{\ell}+b_{\ell}+1)^2}$ \\

        $a_{\ell+1} = b_{\ell} - 1$ & $-a_{\ell}x-\frac{a_{\ell} (b_{\ell}-c_h)}{a_{\ell}-b_{\ell}+1}$ & $a_{\ell}x+\frac{a_{\ell} (b_{\ell}-c_h)}{a_{\ell}-b_{\ell}+1}$ & $- \frac{a_{\ell} (b_{\ell}-1) (-a_{\ell}+c_h-1) (c_h-b_{\ell})}{(a_{\ell}-b_{\ell}+1)^2}$\\
        \hline
    \end{tabular}
    \caption{Expressions of $W_{\ell}^\pm(x)$ and $E_{\ell}(x)$ have been presented for various cases for hypergeometric differential equation supporting a ladder structure for $f_\ell(x)=1$. In all these cases, both $(a_\ell+b_\ell)$ and $c_\ell=c_h$ should be $\ell$-independent.}
    \label{table_Hypgeo_Ladder_2}
\end{table}
In this table, we have quoted the possible ladder structures only for the choices of $\{\Delta(x),p_\ell(x),q_\ell(x),f_\ell(x)\}$ mentioned above. Of course, if one modifies these choices, the corresponding ladder structure (if exists) will change accordingly. For example, one can explicitly check that the hypergeometric differential equation in \ref{hyp} admits a distinct ladder structure if we identify $\Delta(x)=f_\ell(x)=\sqrt{x(x-1)}$, $p_\ell(x) = [(a_\ell+b_\ell+1)x-c_\ell]/\Delta(x)$, and $q_\ell(x) = -a_\ell b_\ell$. In this case, ladder structure exists if $a_\ell+b_\ell=c_1-2\ell$, $c_\ell=c_2-\ell$ with $c_{1,2}$ being constants, and either $a_\ell+a_{\ell-1}=-2\ell+c_1+1$, and/or $a_\ell-a_{\ell-1}=-1$. A concrete illustration of a limiting case of this example in the context of BH spin-ladder is discussed in Ref.~\cite{Hui:2021vcv}. This example also illustrates an important feature of ladder structures. Given a 2nd order OLDE, there are multiple ways to match it to the standard form in \ref{Hl}. And, even if a ladder does not exist for a particular matching, the system may still admit a ladder structure under a different identification.

\subsection{Ladder in confluent hypergeometric differential equation}\label{appB.2}
The confluent hypergeometric equation is given by
\begin{equation} \label{conf}
    x\, \psi_\ell''(x) + (c_\ell-x)\, \psi_\ell'(x) -a_\ell\, \psi_\ell(x) = 0,
\end{equation}
After multiplying by $-x$, it can be written as \ref{Hl}, with
\begin{equation}\label{Confluent_Hypergeometric_Hamiltonian_comarison}
    \Delta(x) = x,\qquad p_{\ell}(x) = c_\ell-x, \qquad q_{\ell}(x) = a_\ell\, x.
\end{equation}
If we consider $f_\ell(x) = 1$, then then the first equation in \ref{conditions} suggests $c_\ell=c_{ch}$ should be $\ell$-independent and the litmus-test condition in \ref{litmus} for the existence of ladder gives
\begin{equation}
    a_{\ell+1}-a_{\ell} = \pm 1,
\end{equation}
The relevant quantities associated with the ladder structure are then summarized in \ref{table_Conf_Hypgeo_Ladder}.

\begin{table}[ht]
    \centering
    \setlength{\tabcolsep}{20pt}
    \renewcommand{\arraystretch}{2}
    \begin{tabular}{cccc}
    \hline
         Cases & $W_{\ell}^+(x)$ & $W_{\ell}^-(x)$ & $E_{\ell}(x)$ \\ \hline
         $a_{\ell+1} = a_{\ell} + 1$ &  $-a_\ell$ & $-x+c_{ch}-a_\ell$ & $ (1-a_\ell) (-c_{ch}+a_\ell)$\\
        
        $a_{\ell+1} = a_{\ell} - 1$ &  $x-c_{ch} + a_\ell$ & $a_\ell$ & $a_\ell (c_{ch}-a_\ell-1)$\\ 
        \hline
    \end{tabular}
    \caption{Expressions of $W_{\ell}^\pm(x)$ and $E_{\ell}(x)$  for various cases for confluent hypergeometric differential equation supporting a ladder structure for $f_\ell(x) = 1$. In both these cases, $c_\ell=c_{ch}$ should be $\ell$-independent.}
    \label{table_Conf_Hypgeo_Ladder}
\end{table}
As discussed for the hypergeometric case, a different identification of $\{\Delta(x),p_\ell(x),q_\ell(x),f_\ell(x)\}$ may lead to a distinct ladder structure from \ref{table_Conf_Hypgeo_Ladder}. Moreover, let us illustrate another important feature of ladder structure using the example of confluent hypergeometric case. We can rewrite the same differential equation into an alternative form by a field redefinition, say, $\psi_\ell(x) = x^{-c_\ell/2}\, e^{x/2}\, \widetilde{\psi}_\ell(x)$. Then, \ref{conf} reduces to a Schr\"odinger-like form  
\begin{equation}
    -\widetilde{\psi}''_\ell(x)+ \left[\frac{\frac{c_\ell}{2}\left(\frac{c_\ell}{2}-1\right)}{x^{2}}-\frac{\frac{c_\ell}{2}-a_\ell}{x}+\frac{1}{4}\right]\widetilde{\psi}_\ell(x)=0.
\end{equation}
Matching with \ref{Hl}, we may now identify $\Delta(x)=1$, $p_\ell(x)=0$, and $q_{\ell}(x)$ being the potential term in the above expression. Then, with $f_\ell(x)=1$, the existence condition for ladder now requires $2(a_{\ell+1}-a_{\ell}) =  c_{\ell+1} - c_{\ell}$, and $c_{\ell+1}=c_{\ell} \pm 2$ and/or $c_{\ell}+c_{\ell + 1} = 4$ and/or $c_\ell+c_{\ell + 1} = 0$. These are starkly different from the earlier condition on $c_\ell$ being $\ell$-independent. Hence, in general, a field redefinition can not only modify the existing ladder structure, but may also lead to situations where a given form of the 2nd order OLDE does not admit a ladder, while the field-redefined equation does. The same note applies to coordinate transformations too.

\twocolumngrid
 
\bibliography{reference_1}

\end{document}